\documentstyle[aps,eqsecnum]{revtex}

\begin{document}

\title{
Formulation for the internal motion of quasi-equilibrium \\ 
configurations in general relativity
} 

\author{
Hideki Asada \footnote{Electronic address: asada@phys.hirosaki-u.ac.jp}\\
Faculty of Science and Technology \\
Hirosaki University, Hirosaki 036-8561, Japan
} 


\maketitle

\begin{abstract}
We present a formulation for the internal motion of equilibrium 
configurations with a rotational Killing vector in general relativity. 
As an approximation, this formulation is applicable to investigation of 
the internal motion of quasi-equilibrium configurations 
such as binary neutron stars. 
Based on this simple formulation, a condition for the general relativistic  
counter rotation has been obtained, though in the recent work 
by Bonazzola, Gourgoulhon and Marck, 
their condition for the counter rotation is not enough to specify 
the internal velocity field. 
Under the condition given in this paper, the internal velocity field 
can be determined completely.  
Indeed, in the counter-rotating case, we have also derived Poisson equations 
for the internal velocity, which take tractable forms 
in numerical implementation. 
\end{abstract}

\begin{flushleft}
PACS Number(s): 04.25.Nx, 04.25.Dm, 04.40.Dg
\end{flushleft}

\def\pa{\partial}


\section{Introduction}

Kilometer-size interferometric gravitational wave detectors, 
such as LIGO\cite{abramovici}, VIRGO\cite{bradaschia} and 
TAMA\cite{kuroda} are now under construction, which 
are increasing our expectation of direct detection 
of gravitational waves. 
Coalescing binary neutron stars are among the most promising sources 
for such detectors. 
The main reasons are that (1) we can expect to detect the signal of 
coalescence of binary neutron stars about several times per year 
\cite{phinney}, and
(2) the waveform from coalescing binaries can be predicted with a high
accuracy compared with other sources \cite{abramovici,thorne94,will94}. 
Informations carried by gravitational waves tell us not only 
various physical parameters of neutron stars\cite{thorne87,thorne94}, 
but also the cosmological parameters\cite{schutz86,mark,finn} 
if and only if we can make a detailed comparison 
between the observed signal with theoretical prediction during 
the epoch of the so-called inspiraling phase where the orbital separation 
is much larger than the radius of component stars \cite{cab}.

To obtain such a large amount of informations of the gravitational
waves, we must understand the theoretical mechanism of merging and 
construct templates for signals a priori. 
When the orbital separation of binary neutron stars is $\leq 10GM/c^2$, 
where $M$ is the total mass of binary neutron stars and $c$ is the light 
velocity, they move approximately in circular orbits 
because the timescale of the energy loss due to gravitational radiation 
$t_{GW}$ is much longer than the orbital period $P$ as 
\begin{equation}
{t_{GW} \over P} \approx 15 \Bigl( {dc^2 \over 10GM} \Bigr)^{5 /2} 
\Bigl( {M \over 4\mu} \Bigr) , 
\label{timescale}
\end{equation}
where $\mu$ and $d$ are the reduced mass and the separation 
of binary neutron stars. 
(As for higher order corrections, see \cite{blanchet} for instance.)
Thus, binary neutron stars evolve adiabatically radiating gravitational 
waves. 
As the orbital separation approaches $6GM/c^2$, however, 
they cannot be described by the circular orbital motion 
because of instabilities due to the general relativistic gravity 
and/or the tidal field. 
As a consequence of such instabilities, binary neutron stars 
will plunge to merge. 
Thus, it is expected that the property of the signal of gravitational waves 
changes drastically around this transition epoch. 
This means that gravitational waves emitted at this transition region 
may bring us an important information about the internal structure 
of neutron stars, since the location where the instability 
occurs, the so-called innermost stable circular orbit (ISCO) 
significantly depends on the equation of state of neutron stars 
\cite{lrs93,lrs94,zcm}. 

A strategy to search the ISCO accurately is as follows:  
Close binary neutron stars have two dominant time scales, 
the orbital period and the time scale of the energy loss. 
The former is much shorter than the latter according to 
Eq.($\ref{timescale}$). 
{}From this physical point of view, we may consider that binary neutron stars 
evolve in the quasi-stationary manner, so that we can take 
the following procedure; 
first, we neglect the effect due to gravitational radiation and 
construct equilibrium configuration, and then the radiation reaction 
is taken into account as a correction to the equilibrium configuration. 
However, how to separate the stationary part from the non-stationary part 
is one of the most important but difficult problems 
in general relativity.  
For instance, Detweiler proposed that a stationary solution of 
the Einstein equation with standing gravitational waves will be 
constructed by adding incoming waves from infinity, and 
may be a valuable approximation to physically realistic solutions 
\cite{d}. 
However, these solutions are not asymptotically flat because 
the total energy of gravitational waves inside a radius $r$ 
grows linearly with $r$. 
Thus, the boundary condition for them is significantly different from 
that of physical solutions. 

In the numerical approach, Wilson and Mathews \cite{wm} proposed 
a semi-relativistic approximate method in order to obtain 
the equilibrium configurations of binary neutron stars just before merging. 
In their method, they assumed the spatially conformal flat metric 
for binary neutron stars. 
Practically, following their approach, numerical calculations have been done 
by some people \cite{wmm,cst,bcsst}. 
However binary neutron stars are genuinely anisotropic, so that 
the assumption of the spatially conformal flatness should 
break down at some level. 
As a result, a lot of debates are continuing \cite{bh,w,f,t}. 

In the post-Newtonian approximation, the metric and the material quantities 
are expanded with respect to $c^{-1}$ assuming the slow motion and 
weak gravity \cite{c65,c67,c69a,cn,ce}. 
In this approximation, we can identify the radiation reaction terms 
which begin at the 2.5PN order \cite{ce}. 
Thus, it is possible to construct the equilibrium configuration 
of binary neutron stars in the 2PN approximation \cite{asf,as}. 
Asada and Shibata presented a formalism to obtain equilibrium configurations 
of uniformly rotating fluid at the 2PN order \cite{as}. 
In this formalism, the equations to be solved are reduced to 
Poisson equations. 
This formulation is applicable to construction of non-axisymmetric uniformly 
rotating equilibrium configurations which include 
synchronized binary neutron stars and the Jacobi ellipsoid. 

Nevertheless, owing to Kochanek's investigation, the viscosity inside 
close binary neutron stars may be significantly small, so that
synchronization may not be a good approximation \cite{k}. 
Rather, irrotation (counter rotation) may be preferred 
for close binary neutron stars. 
Recently, Bonazzola, Gourgoulhon and Marck presented a formulation of 
the general relativistic Euler equation for perfect fluid hydrodynamics 
under the maximal slice \cite{bgm}. 
Especially, in the case of the counter rotation, they gave 
Poisson equations to determine internal velocity fields, 
where they put only one condition on some scalar so that the counter 
rotation could be chosen from various kinds of internal motions 
allowed by the equation of motion. 
However, in order to specify the internal spatial velocity field which 
can be expressed as an essentially three vector, we must adopt 
three conditions.   
Indeed, although one of their equations (Eq.(47) in \cite{bgm}) is 
the Poisson equation for the internal velocity, it turns out to be 
the equation projected perpendicularly to the internal velocity. 
This will be described in detail in section 7. 
Therefore, only their equation is insufficient to determine 
the velocity field. 
The main purpose of this paper is to investigate carefully the equations to 
determine the internal velocity field with respect to the orbital motion, 
and present the equations to determine completely 
the internal velocity field in the case of the general relativistic 
counter rotation. 
In section 2, we assume the rotational Killing vector and introduce 
basic quantities. 
We consider the conservation law with the above point in mind 
in section 3. 
In section 4, the counter-rotating case is considered. 
In section 5, some basic equations are derived, and 
based on this formulation, the condition for the counter rotation 
is presented. 
Next, equations to determine completely the internal velocity field 
are derived in section 6. 
In Section 7, we discuss the relation with this formulation 
with the previous work. 
Section 8 summarizes the conclusions. 
Appendix shows the relation between the counter rotation and the irrotation. 

We use the units of $c=G=1$ in this paper. Greek and Latin indices take 
$0,1,2,3$ and $1,2,3$, respectively.

\section{Killing vector and two frames}

\subsection{Killing vector}

Motivated by the idea discussed in the Introduction, 
let us consider the equilibrium configurations by assuming 
the Killing vector such as 
\begin{equation}
l^{\alpha}=t^{\alpha}+\Omega m^{\alpha} ,   
\end{equation}
where $t^{\alpha}=({\pa \over \pa t})^{\alpha}$ and 
$m^{\alpha}=({\pa \over \pa\phi})^{\alpha}$ . 
If one uses the Cartesian coordinate, $m^{\alpha}$ may be expressed as 
\begin{equation}
m^{\alpha}=-y \Bigl({\pa \over \pa x}\Bigr)^{\alpha}
+x \Bigl({\pa \over \pa y}\Bigr)^{\alpha} . 
\end{equation}
The Killing equation is 
\begin{equation}
{\cal L}_l g_{\alpha\beta}=\nabla_{\alpha}l_{\beta}+\nabla_{\beta}l_{\alpha}
=0 , \label{killing}
\end{equation}
where ${\cal L}$ denotes the Lie derivative. 

We consider a foliation of the spacetime described by the hypersurface 
$(\Sigma_t)$ whose unit normal vector is
\begin{eqnarray}
n_{\mu}&=&(-N,\vec{0}) , \\
n^{\mu}&=&({1 \over N},-{N^i \over N}) , 
\end{eqnarray}
where $N$ and $N^i$ are the lapse function and 
the shift vector respectively \cite{adm,nok}. 
The line element is expressed as 
\begin{equation}
ds^2=-N^2 dt^2+g_{ij}(dx^i+N^i dt)(dx^j+N^j dt) , 
\end{equation}
where our sign convention for the shift vector is opposite to 
that of Bonazzola et.al \cite{bgm}. 
We consider the internal motion which may deviate from the orbital motion 
(co-rotation).  
Therefore, it is convenient to introduce two frames 
for the purpose of the detailed investigation \cite{bgm}; 
one is the non-rotating frame and the other is the co-rotating frame. 

(1) non-rotating frame 

At first, we consider the non-rotating frame as follows: 
The temporal vector $t^{\alpha}$ is related with the unit normal 
vector $n^{\alpha}$ as 
\begin{equation}
t^{\alpha}=N n^{\alpha}+N^{\alpha} .  
\end{equation}

Then, in this frame, we rewrite the Killing vector $l^{\alpha}$ as 
\begin{equation}
l^{\alpha}=N n^{\alpha}+B^{\alpha} , 
\end{equation}  
where we defined $B^{\alpha}$ as 
\begin{equation}
B^{\alpha}=N^{\alpha}+\Omega m^{\alpha} . 
\end{equation}

We define the extrinsic curvature as 
\begin{equation}
K_{\alpha \beta}=-(P_n)^{\mu}_{\alpha}(P_n)^{\nu}_{\beta} 
\nabla_{\mu} n_{\nu} , 
\end{equation}
where the projection tensor $(P_n)_{\alpha}^{\beta}$ was defined as 
\begin{equation}
(P_n)_{\alpha}^{\beta}=\delta_{\alpha}^{\beta}+n_{\alpha}n^{\beta} . 
\end{equation}
Then, we obtain \cite{y}
\begin{equation}
\nabla_{\alpha} n_{\beta}=-K_{\alpha \beta}-n_{\alpha} D_{\beta}\ln{N} , 
\end{equation}
where $D_{\alpha}$ denotes the spatial covariant derivative defined 
on $\Sigma_t$ and we used $n_{\mu}dx^{\mu}=-Ndt$. 

(2) co-rotating frame 

The decomposition with respect to $l^{\alpha}$ is 
useful to describe the internal motion relative to the co-rotation 
along $l^{\alpha}$. 
Let us introduce the unit tangent vector along $l^{\alpha}$ as 
\begin{equation}
v^{\alpha}=e^{-\Phi} l^{\alpha} , 
\label{v}
\end{equation}
and the projection tensor $(P_v)_{\alpha}^{\beta}$ as 
\begin{equation}
(P_v)_{\alpha}^{\beta}=\delta_{\alpha}^{\beta}+v_{\alpha}v^{\beta} . 
\end{equation}
Thus, in the co-rotating frame, we use the projection 
tensor $(P_v)_{\alpha}^{\beta}$, in place of $(P_n)_{\alpha}^{\beta}$ 
in the non-rotating frame. 

{}From Eq.($\ref{v}$), $v_{\alpha}v^{\alpha}=-1$ becomes 
\begin{equation}
e^{2\Phi}=N^2-B_{\alpha}B^{\alpha} . 
\end{equation}
{}From Eq.($\ref{v}$), we obtain \cite{bgm}
\begin{equation}
\nabla_{\beta}v_{\alpha}=\omega_{\alpha\beta}-v_{\beta}\nabla_{\alpha}\Phi , 
\end{equation}
where using the anti-symmetrization $[\;\;]$ we defined 
\begin{equation}
\omega_{\alpha\beta}=-(P_v)_{\alpha}^{\mu}(P_v)_{\beta}^{\nu}
\nabla_{[\mu}v_{\nu]} . 
\label{omega1} 
\end{equation}
Here, introducing 
\begin{equation}
\bar\epsilon^{\alpha\beta\gamma}=v_{\mu}\epsilon^{\mu\alpha\beta\gamma} , 
\end{equation}
we define the vorticity of $v^{\alpha}$ as 
\begin{equation}
\omega^{\alpha}=-{1 \over 2}\bar\epsilon^{\alpha\mu\nu}\omega_{\mu\nu} ,
\label{omega2}
\end{equation}
where we used 
\begin{equation} 
\bar\epsilon_{\alpha\beta\gamma}\bar\epsilon^{\mu\nu\gamma}=
(P_v)^{\mu}_{\alpha}(P_v)^{\nu}_{\beta}
-(P_v)^{\nu}_{\alpha}(P_v)^{\mu}_{\beta} . 
\end{equation}

\section{Conservation Law}

\subsection{adiabatic flow} 

Since we consider the perfect fluid in this paper, 
the stress energy tensor is written as 
\begin{equation}
T^{\mu\nu}=(e+P)u^{\mu}u^{\nu}+P g^{\mu\nu} . 
\end{equation}
We decompose the four velocity in terms of $v^{\alpha}$ as 
\begin{equation}
u^{\mu}=\Gamma (v^{\mu}+V^{\mu}), 
\end{equation}
where we defined $\Gamma$ and $V^{\mu}$ as 
\begin{eqnarray}
\Gamma&=&-v_{\mu}u^{\mu} 
\label{gamma} , \\
V^{\alpha}&=&\Gamma^{-1} (P_v)^{\alpha}_{\mu}u^{\mu} . 
\label{V}
\end{eqnarray}
Then, from $u_{\mu}u^{\mu}=-1$, we obtain 
\begin{equation}
\Gamma=(1-V_{\mu}V^{\mu})^{-1/2} . 
\label{gamma2}
\end{equation}

Next, for the adiabatic flow, the first law of the Thermodynamics leads to 
\begin{equation}
{\nabla_{\alpha} e \over e+P}={\nabla_{\alpha} n \over n} , 
\label{firstlaw}
\end{equation}
where we assumed the barotropic equation of state, $e=e(n)$ and $P=P(n)$. 
Instead of the specific enthalpy 
\begin{equation}
H={e+P \over m_B n} , 
\end{equation}
where $m_B$ is the baryon rest mass, we use the logarithmic enthalpy as 
\begin{equation}
h=\ln{H} . 
\label{h}
\end{equation}
Eq.($\ref{firstlaw}$) is rewritten as 
\begin{equation}
\nabla_{\alpha}h={\nabla_{\alpha}P \over e+P} . 
\end{equation}

\subsection{conservation law}

Here, we consider the conservation law \cite{bgm}. 
{}From its decomposition with respect to $v_{\nu}$, we obtain 
\begin{equation}
\nabla_{\mu}V^{\mu}+V^{\mu}\nabla_{\mu} 
\Bigl( \ln{(\Gamma^2 n)}+\Phi+h \Bigr) , 
\label{cons}
\end{equation}
and 
\begin{eqnarray}
\nabla_{\mu}(V^{\mu}V^{\alpha})+2\omega^{\alpha\mu}V_{\mu}
+\nabla^{\alpha}\Phi+\Gamma^{-2}\nabla^{\alpha}h&& 
\nonumber\\\
+V^{\alpha}V^{\mu}\nabla_{\mu} \Bigl( \ln{(\Gamma n)}-\Phi \Bigr) &=&0 . 
\label{vertcons}
\end{eqnarray}

Moreover, by decomposing Eq.($\ref{vertcons}$) with respect to $V^{\alpha}$, 
we obtain 
\begin{equation}
V^{\mu}\nabla_{\mu}(\ln{\Gamma}+\Phi+h)=0 , 
\label{bernoulli}
\end{equation}
which can be taken as the expression for a general relativistic version of 
the Bernoulli theorem. 
Namely, $\ln{\Gamma}+\Phi+h$ is constant along the internal flow 
$V^{\alpha}$. 
Then, Eq.($\ref{cons}$) becomes 
\begin{equation}
\nabla_{\mu}V^{\mu}+V^{\mu}\nabla_{\mu}\ln{(n\Gamma)}=0 , 
\label{conti}
\end{equation}
which is exactly the equation for the baryon number conservation. 
Furthermore, we obtain the general relativistic analogue of the Euler's 
equation as 
\begin{equation}
V^{\mu}\nabla_{\mu}V^{\alpha}+2\bar\epsilon^{\alpha\mu\nu} 
\omega_{\mu}V_{\nu}+\nabla^{\alpha}\Phi+\Gamma^{-2}\nabla^{\alpha}h
-V^{\alpha}V^{\mu}\nabla_{\mu}\Phi=0 , 
\label{euler1}
\end{equation}
where we used Eq.($\ref{bernoulli}$). 

In order to obtain simple expressions later, it will be useful to introduce 
the vorticity of the internal motion as 
\begin{equation}
\Omega^{\alpha}={1 \over 2}\bar\epsilon^{\alpha\mu\nu}\nabla_{\mu}V_{\nu} . 
\end{equation}
Using the internal vorticity, we obtain the relation 
\begin{equation}
V^{\mu}\nabla_{\mu}V^{\alpha}=\Gamma^{-2}\nabla^{\alpha}\ln{\Gamma}
+2\bar\epsilon^{\alpha\mu\nu}\Omega_{\mu}V_{\nu} , 
\end{equation}
where we used Eq.($\ref{gamma2}$). 
Using this relation, Eq.($\ref{euler1}$) becomes  
\begin{equation}
\nabla^{\alpha}(\ln{\Gamma}+\Phi+h)+\Gamma^2 \Bigl( 
2\bar\epsilon^{\alpha\mu\nu}(\Omega_{\mu}+\omega_{\mu})V_{\nu}
+V_{\mu}V^{\mu}(P_V)^{\alpha}_{\nu}\nabla^{\nu}\Phi \Bigr)=0 , 
\label{euler2}
\end{equation}
where we defined the projection operator with respect to 
the internal velocity as 
\begin{equation}
(P_V)_{\alpha}^{\beta}=\delta_{\alpha}^{\beta}
-{V_{\alpha}V^{\beta} \over V_{\mu}V^{\mu}} . 
\end{equation}

\section{Counter-rotation} 

In order to simplify the expression, we introduce the scalar 
${\cal G}$ as 
\begin{equation}
{\cal G}=\ln{\Gamma}+\Phi+h . 
\label{G}
\end{equation}
This can be re-expressed as 
\begin{equation}
{\cal G}=\ln{\Bigl(-u_{\nu}l^{\nu} H\Bigr)} . 
\end{equation}
Then, Eqs.($\ref{bernoulli}$) and ($\ref{euler2}$) become respectively 
\begin{equation}
V^{\mu}\nabla_{\mu}{\cal G}=0 ,  
\label{bernoulli2}
\end{equation}
and 
\begin{equation}
2 \bar\epsilon_{\alpha\mu\nu}(\Omega^{\mu}+\omega^{\mu})V^{\nu}
+V_{\mu}V^{\mu}(P_V)^{\nu}_{\alpha}\nabla_{\nu}\Phi
+\Gamma^{-2}\nabla_{\alpha}{\cal G}=0 . 
\label{euler3}
\end{equation}
It should be noted that we do not need to solve Eq.($\ref{bernoulli2}$), 
since it is obtained by contracting Eq.($\ref{euler3}$) 
with $V^{\alpha}$. 

In the Newtonian limit, the counter rotation means 
\begin{equation}
\Omega_{\mu}+\omega_{\mu} \to 0 . 
\end{equation}
Therefore, a simple condition for the counter rotation would be 
\begin{equation}
\Omega_{\mu}+\omega_{\mu}=0 . 
\end{equation}
In this case, however, Eq.($\ref{euler3}$) becomes 
\begin{equation}
\Gamma^2 V_{\mu}V^{\mu}(P_V)^{\nu}_{\alpha}\nabla_{\nu}\Phi
+\nabla_{\alpha}{\cal G}=0 ,  
\end{equation}
for which such a ${\cal G}$ does not always exist. 

It seems difficult to find out conditions for the internal motion 
in the counter-rotating case from Eq.($\ref{euler3}$).  
Therefore, it is desirable to change Eq.($\ref{euler3}$) 
into the form easier to investigate. 
Indeed, we can reformulate the equation as follows: 
First, we should note the relation 
\begin{equation}
\Omega^{\alpha}+{1 \over 2}\bar\epsilon^{\alpha\mu\nu}V_{\mu}\nabla_{\nu}\Phi
=\tilde\Omega^{\alpha} ,
\end{equation}
where we defined the {\it renormalized} vorticity as 
\begin{equation}
\tilde\Omega^{\alpha}={1 \over 2} e^{\Phi}\bar\epsilon^{\alpha\mu\nu}
\nabla_{\mu}(e^{-\Phi}V_{\nu}) . 
\label{renvol}
\end{equation}
Then, we obtain  
\begin{equation}
\bar\epsilon_{\alpha\mu\nu}\Omega^{\mu}V^{\nu}
+{1 \over 2}V_{\nu}V^{\nu}(P_V)^{\beta}_{\alpha}\nabla_{\beta}\Phi
=\bar\epsilon_{\alpha\mu\nu}\tilde\Omega^{\mu}V^{\nu} . 
\end{equation}
Using this relation, we rewrite Eq.($\ref{euler3}$) in the simple form as  
\begin{equation}
\bar\epsilon_{\alpha\mu\nu}(\tilde\Omega^{\mu}+\omega^{\mu})V^{\nu}
+{1 \over 2}\Gamma^{-2}\nabla_{\alpha}{\cal G}=0 . 
\label{euler4}
\end{equation}

Owing to the simple form of Eq.($\ref{euler4}$), 
we can define the counter rotation by requiring 
$\tilde\Omega^{\nu}$ such as 
\begin{equation}
\tilde\Omega^{\nu}+\omega^{\nu}=0 . 
\label{counter}
\end{equation}
In the Newtonian limit, this coincides with the condition 
for the Newtonian counter rotation. 
Under Eq.($\ref{counter}$), Eq.($\ref{euler4}$) means simply  
\begin{equation}
{\cal G}=\mbox{constant} ,
\label{gconst}
\end{equation}
which satisfies a general relativistic version of the 
Bernoulli theorem ($\ref{bernoulli}$). 
It is noteworthy that the condition $(\ref{counter})$ satisfies 
the irrotation condition. (For instance, see \cite{he} for the definition 
of the vorticity). This is shown in the appendix. 

\section{Basic equations for internal velocity field}
\subsection{decomposition of $V^{\alpha}$}

In the previous section, we have presented Eq.($\ref{counter}$), 
the condition for the general relativistic version of the counter rotation,  
by the use of Eq.($\ref{euler3}$) which is an algebraic equation 
for the vorticity $\Omega^{\alpha}$. 
For the numerical implementation, 
it is desirable to derive the equations for the internal velocity field 
in the case of the counter rotation. 
First, with respect to $\Sigma_t$, we decompose the internal velocity 
$V^{\alpha}$ into the following form 
\begin{equation}
V^{\alpha}=V_{\bot}^{\alpha}+V_{\Vert}^{\alpha} , 
\end{equation}
where we defined 
\begin{eqnarray}
V_{\bot}^{\alpha}&=&(P_n)^{\alpha}_{\beta}V^{\beta} , \\
V_{\Vert}^{\alpha}&=&V_{\Vert}n^{\alpha} , \\
V_{\Vert}&=&-n_{\beta}V^{\beta} . 
\end{eqnarray}

Since the spatial vector $V^{\alpha}$ expresses only three degrees 
of freedom in the four velocity, $V_{\Vert}$ is related with 
$V_{\bot}^{\alpha}$. 
Indeed, from $v_{\alpha}V^{\alpha}=0$, we obtain  
\begin{equation}
V_{\Vert}={1 \over N}B_iV^i_{\bot} . 
\end{equation}
For the sake of the further decomposition, we introduce the spatial 
alternating tensor as 
\begin{equation}
\hat\epsilon^{\alpha\beta\gamma}=n_{\mu}\epsilon^{\mu\alpha\beta\gamma} . 
\end{equation}
This new alternating tensor is related with 
$\bar\epsilon^{\alpha\beta\gamma}$ as 
\begin{equation}
\bar\epsilon^{\alpha\beta\gamma}=e^{-\Phi} \Bigl\{ 
N\hat\epsilon^{\alpha\beta\gamma} +B_{\mu} \Bigl( 
n^{\alpha}\hat\epsilon^{\mu\beta\gamma}
+n^{\beta}\hat\epsilon^{\mu\gamma\alpha}
+n^{\gamma}\hat\epsilon^{\mu\alpha\beta} \Bigr) \Bigr\} . 
\label{alternating}
\end{equation}
In turn, with respect to $n^{\alpha}$, we decompose the covariant 
derivative of the internal velocity as 
\begin{eqnarray}
\nabla_{\alpha}V_{\beta}&=&D_{\alpha}V_{\bot\, \beta}+{n_{\alpha} \over N}
[B,V_{\bot}]_{\beta}+(n_{\alpha} K_{\beta\gamma}-n_{\beta} K_{\alpha\gamma}) 
V_{\bot}^{\gamma}-n_{\alpha}n_{\beta}V_{\bot}^{\gamma}D_{\gamma}\ln{N}
\nonumber\\
&&+n_{\beta}\nabla_{\alpha}V_{\Vert}-V_{\Vert}(K_{\alpha\beta}
+n_{\alpha}D_{\beta}\ln{N}) , 
\label{gradV}
\end{eqnarray}
where we defined the Lie bracket as 
\begin{equation}
[B,V_{\bot}]^{\alpha}=B^{\beta}\nabla_{\beta}V_{\bot}^{\alpha}
-V_{\bot}^{\beta}\nabla_{\beta}B^{\alpha} . 
\end{equation}
Here, we used the following relation for the vector $X^{\alpha}$ 
on $\Sigma_t$ 
\begin{equation}
n^{\beta}\nabla_{\beta}X^{\alpha}=n^{\alpha}X^{\beta}D_{\beta}\ln{N}
-X_{\beta}K^{\alpha\beta}-{1 \over N}[B,X]^{\alpha} , 
\label{lie}
\end{equation}
which is derived from 
\begin{equation}
{\cal L}_l X^{\alpha}=0 . 
\end{equation}
Therefore, by the frequent use of Eq.($\ref{lie}$) in Eq.($\ref{renvol}$), 
we obtain 
\begin{eqnarray}
\tilde\Omega^{\alpha}&=&{1 \over 2}e^{-\Phi}\hat\epsilon^{\alpha\mu\nu} 
\Bigl[ N D_{\mu}V_{\bot\, \nu}+B_{\mu} \Bigl\{ -{1 \over N} 
[B,V_{\bot}]_{\bot\, \nu}-2K_{\nu\sigma}V_{\bot}^{\sigma}
-V_{\Vert}D_{\nu}\ln{N}+D_{\nu}V_{\Vert} \Bigr\} \nonumber\\
&&+\Bigl( NV_{\bot\,\mu}-V_{\Vert}B_{\mu} \Bigr)
-{1 \over N}V_{\bot\,\mu}B_{\nu}B^{\sigma}D_{\sigma}\Phi \Bigr]
+{1 \over 2}e^{-\Phi}n^{\alpha}\hat\epsilon^{\lambda\mu\nu} B_{\lambda} 
\Bigl( D_{\mu}V_{\bot\,\nu}+V_{\bot\,\mu}D_{\nu}\Phi \Bigr) , 
\label{Omega2}
\end{eqnarray}
where we defined the Lie bracket on $\Sigma_t$ as 
\begin{equation}
[B,V_{\bot}]_{\bot\,\nu}=B^{\mu}D_{\mu}V_{\bot\,\nu}
-V_{\bot}^{\mu}D_{\mu}B_{\nu}. 
\end{equation}
{}From Eqs.($\ref{omega1}$), ($\ref{omega2}$) and ($\ref{alternating}$), 
we obtain 
\begin{eqnarray}
\omega^{\alpha}&=&{1 \over 2}e^{-2\Phi}\hat\epsilon^{\alpha\mu\nu} 
\Bigl( N D_{\mu}B_{\nu}+2 B_{\mu} D_{\nu}N
+2 K_{\mu}^{\sigma}B_{\sigma}B_{\nu} \Bigr) 
+{1 \over 2}e^{-2\Phi}n^{\alpha}B_{\sigma}\hat\epsilon^{\sigma\mu\nu}
D_{\mu}B_{\nu} . 
\label{omega3}
\end{eqnarray}
{}From Eqs.($\ref{Omega2}$) and ($\ref{omega3}$), Eq.($\ref{counter}$) 
becomes 
\begin{eqnarray}
&&\hat\epsilon^{\alpha\mu\nu} \Bigl[ ND_{\mu} V_{\bot\,\nu}+B_{\mu} \Bigl( 
-{1 \over N} [B,V_{\bot}]_{\bot\,\nu}-2K_{\nu\sigma}V_{\bot}^{\sigma}
-V_{\Vert}D_{\nu} \ln{N}+D_{\nu} V_{\Vert} \Bigr) \nonumber\\
&&~~~~~~~+\Bigl( N V_{\bot \mu}-V_{\Vert} B_{\mu} \Bigr) D_{\nu}\Phi
-{1 \over N}V_{\bot\,\mu}B_{\nu} B^{\sigma} D_{\sigma} \Phi \nonumber\\
&&~~~~~~~+e^{-\Phi} \Bigl( N D_{\mu} B_{\nu}+2B_{\mu} D_{\nu} N
+2K_{\mu}^{\sigma} B_{\sigma} B_{\nu} \Bigr) \Bigr] \nonumber\\
&&+n^{\alpha} \hat\epsilon^{\lambda\mu\nu} B_{\lambda} \Bigl( D_{\mu} 
V_{\bot\,\nu}+V_{\bot\,\mu} D_{\nu} \Phi
+e^{-\Phi} D_{\mu} B_{\nu} \Bigr) =0 . 
\label{counter2}
\end{eqnarray}
This equation has been already decomposed manifestly 
with respect to $n^{\alpha}$. 
Therefore, using the spatial indices, equations to be solved 
can be rewritten as 
\begin{eqnarray}
\hat\epsilon^{ijk} \Bigl[ && ND_j V_{\bot\,k}+B_j \Bigl( 
-{1 \over N} [B,V_{\bot}]_{\bot\,k}-2K_{kl}V_{\bot}^l
-V_{\Vert}D_k \ln{N}+D_k V_{\Vert} \Bigr) \nonumber\\
&&+\Bigl( N V_{\bot j}-V_{\Vert} B_j \Bigr) D_k\Phi
-{1 \over N}V_{\bot\,j}B_k B^l D_l \Phi \nonumber\\
&&+e^{-\Phi} \Bigl( N D_j B_k+2B_j D_k N+2K_j^l B_l B_k \Bigr) \Bigr] =0 , 
\label{counter3}
\end{eqnarray}
and 
\begin{equation}
\hat\epsilon^{ijk} B_i \Bigl( D_j V_{\bot\,k}+V_{\bot\,j} D_k \Phi
+e^{-\Phi} D_j B_k \Bigr) =0 . 
\end{equation}
It should be noted that we need not solve the latter equation, 
since it can be derived from the projection of the former 
onto $B_i$. 
         
By taking the contraction of Eq.($\ref{gradV}$), we obtain 
\begin{equation}
\nabla_{\alpha}V^{\alpha}=D_{\alpha}V_{\bot\,}^{\alpha}
+V_{\bot}^{\alpha}D_{\alpha}\ln{N}-{B^{\alpha} \over N}D_{\alpha}V_{\Vert}
-V_{\Vert} K ,  
\end{equation}
where we used the following identity for arbitrary vectors 
$E^{\alpha}$ and $F^{\alpha}$ on $\Sigma_t$ 
\begin{equation}
n^{\alpha}[E,F]_{\alpha}=0 . 
\end{equation}
Thus, with the spatial indices, Eq.($\ref{conti}$) becomes 
\begin{equation}
D_i V_{\bot}^i +V_{\bot}^i D_i \ln{(Nn\Gamma)}-{V_{\Vert} \over N} 
B^i D_i \ln{(V_{\Vert}n\Gamma)}-V_{\Vert}K=0 . 
\label{conti2}
\end{equation}

Since we have derived the basic equations ($\ref{counter3}$) and 
($\ref{conti2}$) to be solved, all we must do is to investigate 
the boundary condition for these equations. 
Here, we concentrate our attention on conditions 
on the surface of the matter. 
By multiplying Eq.($\ref{conti2}$) with $n$ and evaluating it 
on the surface of the matter, we obtain 
\begin{equation}
\Bigl( V_{\bot}^i-{B^i B^j V_{\bot\,j} \over N^2} \Bigr) (D_i n)  
\Bigr\vert_{\mbox{surface}}=0 , 
\end{equation}
where we used the fact that $n$ must vanish on the surface. 
This is the boundary condition for Eqs.($\ref{counter3}$) 
and ($\ref{conti2}$). 

\section{Poisson equations for the internal velocity field}

In order to reduce the problem of solving Eqs.($\ref{counter3}$) 
and ($\ref{conti2}$) to that of solving the Poisson equations, we decompose 
$V_{\bot}^i$ as follows; 
\begin{equation}
V_{\bot}^i =\hat\epsilon^{ijk} D_j A_k +D^i \psi , 
\end{equation}
where in order to fix gauge freedoms in this decomposition, 
we adopt the Coulomb gauge 
\begin{equation}
D_i A^i =0 . 
\end{equation}
Applying this decomposition to Eq.($\ref{counter3}$), we obtain
\begin{eqnarray}
D_jD^j A^i &=&{}^3R^i_k A^k \nonumber\\
&&+\hat\epsilon^{ijk} \Bigl[ -B_j \Bigl( {1 \over N^2}[B,V_{\bot}]_{\bot\,k}
+{2 \over N}K_{kl}V_{\bot}^l+{V_{\Vert} \over N^2} D_k N 
-{1 \over N} D_k V_{\Vert} \Bigr) \nonumber\\
&&~~~~~~~~+\Bigl( V_{\bot j}-{V_{\Vert} \over N}B_j \Bigr) D_k \Phi 
-{1 \over N^2}V_{\bot\,j}B_k B^l D_l \Phi \nonumber\\
&&~~~~~~~~+e^{-\Phi} \Bigl( D_j B_k+{2 \over N} B_j D_k N
+{2 \over N}K_j^l B_l B_k \Bigr) \Bigr] , 
\label{eqA}
\end{eqnarray}
where ${}^3R^i_k$ denotes the Ricci tensor defined on $\Sigma_t$ and 
we used 
\begin{equation}
\hat\epsilon^{ijk} D_j V_{\bot\,k}=-D_jD^j A^i+{}^3R^i_k A^k . 
\end{equation}
Similarly, from Eq.($\ref{conti2}$), we obtain 
\begin{equation}
D_jD^j \psi=-V_{\bot}^i D_i \ln{(Nn\Gamma)}+{V_{\Vert} \over N} 
B^i D_i \ln{(V_{\Vert}n\Gamma)}+V_{\Vert}K .  
\label{eqpsi}
\end{equation}

\section{Discussion}

Bonazzola et.al. \cite{bgm} have recently developed the formulation for 
the internal motion of quasi-equilibrium configurations. 
However, they put only one condition for a scalar, so that 
the counter rotation could be chosen from various kinds of internal motions. 
Namely, in our notation, their condition is expressed as 
\begin{equation}
{\cal G}=\mbox{constant} ,  
\label{bgmcounter}
\end{equation}
which is not enough for full specifying the internal vorticity.   
We can understand the reason by counting the degrees of freedom:  
Although the spatial velocity has three degrees of freedom, 
Eq.($\ref{bgmcounter}$) fixes only one degree of freedom. 

We can also realize the situation in the following concrete form: 
The equation (Eq.(50) in \cite{bgm}), which was claimed to determine 
the internal velocity field, is rewritten as 
\begin{equation}
(P_V)^{\alpha}_{\beta}(\Omega^{\beta}+\omega^{\alpha})=-{1 \over 2} 
\bar\epsilon^{\alpha\mu\nu}V_{\mu}\nabla_{\nu}\Phi , 
\label{bgmeq}
\end{equation}
which is obtained by contracting Eq.($\ref{euler3}$) for 
${\cal G}=\mbox{constant}$ with $\bar\epsilon^{\mu\nu\alpha}V_{\nu}$. 
Since this manipulation is the projection perpendicular to $V^{\alpha}$,  
the resultant equation ($\ref{bgmeq}$) shows clearly the projection, 
though in the original form (Eq.(50) in \cite{bgm}) 
the projection of $\Omega^{\alpha}$ is separated into two parts, 
one in the left hand side and the other in the right hand side. 
Therefore, this equation is not enough to determine fully 
the internal velocity field. 
On the other hand, in this paper, we have defined the counter rotation 
by requiring Eq.($\ref{counter}$), so that the full determination of 
the internal motion can be achieved by Eqs.($\ref{eqA}$) 
and ($\ref{eqpsi}$).

\section{Conclusion}

In this paper, we have presented a formulation for the internal motion of 
equilibrium configurations with a rotational Killing vector 
in general relativity. 
As an approximation, this formulation is applicable to investigation of 
the internal motion of quasi-stationary configurations 
such as binary neutron stars. 
In particular, by careful treatment, we have derived the simple equation 
($\ref{euler4}$) to determine the internal velocity relative to 
the orbital motion. 
Thus, using this simple form of the equation, we could find easily 
a condition ($\ref{counter}$) for the general relativistic 
counter rotation. 
This condition determines completely the internal velocity field.  
In this counter-rotating case, we have also presented Poisson equations 
for the internal velocity, which take tractable forms 
in numerical implementation. 
For instance, an iterative procedure has been proposed \cite{bgm}. 
Thus, the formalism presented here will be useful to investigate 
quasi-equilibrium configurations for counter-rotating 
binary neutron stars and/or the Riemann ellipsoid. 
In addition, it is noteworthy that slice conditions are not specified 
in the present paper. 

We focused our attention on the conservation law, 
which gives us the equation of motion. 
In order to construct quasi-equilibrium configurations, we must solve 
the Einstein equation. 
The post-Newtonian approximation is useful to find out 
the stationary parts in the Einstein equation. 
In the counter-rotating case, the equations to determine the metric 
will be derived in the second post-Newtonian approximation, 
in the similar manner in the synchronized case \cite{as}. 
This will be done in the future. 

\acknowledgements 

We would like to thank T. Nakamura, M. Shibata and M. Kasai
for useful conversation.

\appendix
\section{counter rotation and irrotation}
Here, we shall show that the counter rotation $(\ref{counter})$ 
means the irrotation (vanishing vorticity), in the spacetime 
with the rotational Killing vector. 
The vorticity is defined as (for instance, \cite{he})
\begin{equation}
{\bf\Omega}_{\alpha\beta}=-P^{\mu}_{\alpha}P^{\nu}_{\beta} 
\nabla_{[\mu}u_{\nu]} . 
\label{vorticity}
\end{equation}
For the fluid with the barotropic equation of state, the vorticity is 
rewritten as 
\begin{equation}
{\bf\Omega}_{\alpha\beta}=-{1 \over H} \nabla_{[\alpha} (Hu_{\beta]} ) , 
\label{vorticity2}
\end{equation}
where we used the projection of the conservation law with respect to 
$u^{\alpha}$ \cite{bgm} 
\begin{equation}
u^{\mu}\nabla_{\mu}u^{\alpha}+\nabla^{\alpha}h
+u^{\alpha}u^{\mu}\nabla_{\mu}h=0 . 
\end{equation}

{}From Eqs. $(\ref{omega2})$ and $(\ref{renvol})$, we obtain 
\begin{equation}
\tilde\Omega_{\mu}+\omega_{\mu}={1 \over 2} e^{\Phi} 
\bar\epsilon_{\mu\rho\sigma} \nabla^{\rho}\Bigl( e^{-\Phi} 
(V^{\sigma}+v^{\sigma}) \Bigr) .  
\label{counter2a}
\end{equation}
{}From Eqs.$(\ref{h})$, $(\ref{G})$ and $(\ref{gconst})$, we can take 
$e^{-\Phi}$ as 
\begin{equation}
e^{-\Phi}=C \Gamma H , 
\end{equation}
where $C$ is some constant. 
Then, Eq.$(\ref{counter2a})$ becomes 
\begin{equation}
\tilde\Omega_{\mu}+\omega_{\mu}={1 \over 2} C e^{\Phi} 
\bar\epsilon_{\mu\rho\sigma} \nabla^{\rho} (Hu^{\sigma}) . 
\label{counter3a}
\end{equation} 

Inserting Eq.$(\ref{vorticity2})$ into Eq.$(\ref{counter3a})$, 
we obtain  
\begin{equation}
\tilde\Omega_{\mu}+\omega_{\mu}=-{1 \over 2} CH e^{\Phi} 
\bar\epsilon_{\mu\rho\sigma} {\bf\Omega}^{\rho\sigma} . 
\label{counter4a}
\end{equation}
Thus, we find  
\begin{equation}
\epsilon_{\mu\alpha\beta\gamma} ( \tilde\Omega^{\mu}+\omega^{\mu} ) 
=-2CHe^{\Phi} (v_{\alpha}{\bf\Omega}_{\beta\gamma}
+v_{\beta}{\bf\Omega}_{\gamma\alpha}+v_{\gamma}{\bf\Omega}_{\alpha\beta}) . 
\end{equation}
Since the left hand side vanishes in the counter rotation, we obtain  
\begin{equation}
v_{\alpha}{\bf\Omega}_{\beta\gamma}+v_{\beta}{\bf\Omega}_{\gamma\alpha}
+v_{\gamma}{\bf\Omega}_{\alpha\beta}=0 . 
\label{vomega}
\end{equation}
Contracting $u^{\beta}v^{\gamma}$ with this, we find 
\begin{equation}
{\bf\Omega}_{\alpha\gamma}v^{\gamma}=0 . 
\end{equation}
Hence, contracting $v^{\gamma}$ with Eq.$(\ref{vomega})$, we obtain 
\begin{equation}
{\bf\Omega}_{\alpha\beta}=0 . 
\end{equation}


\begin{thebibliography}{999}

\bibitem{abramovici}A. Abramovici et al. Science, {\bf 256}, 325 (1992). 
\bibitem{bradaschia}C. Bradaschia et al., Nucl. Instrum. Method Phys. 
Res. Sect. {\bf A289}, 518 (1990). 
\bibitem{kuroda}K. Kuroda et. al., In proceedings of the international 
conference on gravitational waves: Sources and Detections, 
ed. I. Ciufolini and F. Fidecard (World Scientific, 1997), p. 100. 
\bibitem{phinney}E. S. Phinney, Astrophys. J. Lett. {\bf 380}, 17 (1991). 
\bibitem{thorne94}K. S. Thorne, In proceedings of the eighth 
Nishinomiya-Yukawa memorial symposium on {\it Relativistic Cosmology}, 
ed. M. Sasaki (Universal Academy Press, Tokyo, 1994). 
\bibitem{will94}C. M. Will, In proceedings of the eighth Nishinomiya-Yukawa
memorial symposium on {\it Relativistic Cosmology}, 
ed. M. Sasaki (Universal Academy Press, Tokyo, 1994). 
\bibitem{thorne87}K. S. Thorne, In {\it 300 Years of Gravitation}, 
ed. S. Hawking and W. Israel (Cambridge; 1987) p. 330. 
\bibitem{schutz86}B. F. Schutz, Nature, {\bf 323}, 210 (1986). 
\bibitem{mark}D. Markovic, Phys. Rev. {\bf D48}, 4738 (1993). 
\bibitem{finn}L. S. Finn, Phys. Rev. {\bf D53}, 2878 (1996). 
\bibitem{cab}C. Cutler, T. A. Apostolatos, L. Bildsten, L. S. Finn, 
E. E. Flanagan, D. Kennefick, D. M. Markovic, A. Ori, E. Poisson, 
G. J. Sussman, K. S. Thorne, Phys. Rev. Lett. {\bf 70}, 2984 (1993). 
\bibitem{blanchet}L. Blanchet, Phys. Rev. {\bf D54}, 1417 (1996). 
\bibitem{lrs93}D. Lai, F. Rasio and S. L. Shapiro, Astrophys. J. Suppl. 
{\bf 88}, 205 (1993).
\bibitem{lrs94}D. Lai, F. Rasio and S. L. Shapiro, Astrophys. J. 
{\bf 420}, 811 (1994). 
\bibitem{zcm}X. Zhung, J. M. Centrella, and S. L. W. McMillan, 
Phys. Rev. {\bf D50}, 6247 (1994). 
\bibitem{d}S. Detweiler, Phys. Rev. {\bf D50}, 4929 (1994). 
\bibitem{wm}J. R. Wilson and G. J. Mathews, Phys. Rev. Lett. 
{\bf 75}, 4161 (1995). 
\bibitem{wmm}J. R. Wilson, G. J. Mathews and P. Marronetti, Phys. Rev. 
{\bf D54}, 1317 (1996). 
\bibitem{cst}G. Cook, S. L. Shapiro and S. A. Teukolsky, 
Phys. Rev. {\bf D53}, 5533 (1996). 
\bibitem{bcsst}T. W. Baumgarte, G. B. Cook, M. A. Scheel, S. L. Shapiro 
and S. A. Teukolsky, Phys. Rev. Lett. {\bf 79}, 1182 (1997). 
\bibitem{bh}P. R. Brady and S. A. Hughes, Phys. Rev. Lett. 
{\bf 79}, 1186 (1997). 
\bibitem{w}A. G. Wiseman, Phys. Rev. Lett. {\bf 79}, 1189 (1997). 
\bibitem{f}E. E. Flanagan, submitted to Phys. Rev. D, (gr-qc/9706045). 
\bibitem{t}K. S. Thorne, submitted to Phys. Rev. D, (gr-qc/9706057). 
\bibitem{c65}S. Chandrasekhar, Astrophys. J. {\bf 142}, 1488 (1965). 
\bibitem{c67}S. Chandrasekhar, Astrophys. J. {\bf 148}, 621 (1967).
\bibitem{c69a}S. Chandrasekhar, Astrophys. J. {\bf 158}, 45 (1969a). 
\bibitem{cn}S. Chandrasekhar and Y. Nutku, Astrophys. J. 
{\bf 158}, 55 (1969). 
\bibitem{ce}S. Chandrasekhar and F. P. Esposito, Astrophys. J. 
{\bf 160}, 153 (1970). 
\bibitem{asf}H. Asada, M. Shibata and T. Futamase, 
Prog. Theor. Phys. {\bf 96}, 81 (1996). 
\bibitem{as}H. Asada and M. Shibata, 
Phys. Rev. {\bf D54}, 4944 (1996). 
\bibitem{k}C. S. Kochanek, Astrophys. J. {\bf 398}, 234 (1992). 
\bibitem{bgm}S. Bonazzola, E. Gourgoulhon and J. Marck, 
Phys. Rev. {\bf D56}, 7740 (1997). 
\bibitem{adm}R. Arnowitt, S. Deser and C. W. Misner, In {\it An Introduction 
to Current Research}, ed. L. Witten, (Wiley, New York, 1962). 
\bibitem{nok}T. Nakamura, K. Oohara and Y. Kojima, Prog. Theor. Phys. Suppl. 
{\bf 90} (1987).
\bibitem{y}J. W. York, In {\it Sources of Gravitational Radiation}, 
ed. L. Smarr (Cambridge University Press, Cambridge, 1979). 
\bibitem{he} S. W. Hawking and G. F. R. Ellis, {\it The Large Scale Structure 
of Space-time}, (Cambridge University Press, Cambridge, 1973). 

\end{thebibliography}
\end{document}